\documentclass[showpacs,preprint,aps]{revtex4}
\usepackage{graphicx}

\begin{document}
\begin{center}
{\bf \Large  On Quasibound N$^*$ Nuclei}
\vskip 5mm
N. G. Kelkar, D. Bedoya Fierro\\
Departamento de Fisica, Universidad de los Andes, Cra 1E, 18A-10, 
Bogot\'a, Colombia\\
\vskip 5mm
P. Moskal\\
M. Smoluchowski Institute of Physics, Jagiellonian University, 30-348 Cracow, 
Poland
\end{center}
\begin{abstract}
The possibility for the existence of
unstable bound states of the S11 nucleon resonance N$^*$(1535) and nuclei is 
investigated. These quasibound states are 
speculated to be closely related to the existence of the quasibound states 
of the eta mesons and nuclei. Within a simple model for the 
N N$^*$ interaction involving a pion and eta meson 
exchange, N$^*$-nucleus potentials for N*-$^3$He and N*-$^{24}$Mg are evaluated 
and found to be of a Woods-Saxon like form which supports two to three bound states. 
In case of N*-$^3$He, one state bound by only a few keV and another 
by 4 MeV is found. The results are however quite sensitive to the N N$^*$ $\pi$ and 
N N$^*$ $\eta$ vertex parameters. A rough estimate of the width of these states, based on 
the mean free path of the exchanged mesons in the nuclei leads to very broad states 
with $\Gamma \sim$ 80 and 110 MeV for N*-$^3$He and N*-$^{24}$Mg respectively. 
\end{abstract}
\maketitle
\section{Introduction}
The S11 nucleon resonance N*(1535) has always been considered a crucial ingredient
in the search for the elusive eta mesic nuclei \cite{etamesic1, etamesic2}.
Analyses of an anticipated eta mesic nucleus picture the eta-nucleon interaction to
proceed via the formation of an N*(1535) resonance which repeatedly decays,
regenerates and propagates within the nucleus until it eventually decays into a free
meson and nucleon. 
Such a picture makes one ponder if a quasibound state of the N* and
nucleus might also exist. The idea of a ``bound" state of an N$^*$ and a nucleus is 
conceptually similar to that of a $\Delta$ and a nucleus which was indeed investigated 
in the past. In an experiment performed at MAMI \cite{bartsch}, 
the reaction $^{12}$C($e$,$e^{\prime}\, 
\Delta^0$)$^{11}$C $\to$ $^{12}$C($e$,$e^{\prime}\,p \pi$ )$^{11}$C was
investigated and the authors claimed to have found evidence for two narrow peaks
which they interpreted as $^{12}$C$_{\Delta}$ states.
The authors distinguished the reaction with two scenarios: (i) a
``quasifree" $\Delta^0$ is produced from a bound neutron and it flies off in the
forward direction and decays such that the decay particles are produced in the
forward direction in the laboratory frame and (ii) a bound $\Delta^0$ is produced
and the whole nucleus takes the momentum transfer such that the $\Delta^0$ moves much
slower and the decay products can in principle come out in any direction.
The forward direction decay products were then excluded in order to look for the
bound $\Delta^0$.
Though the authors did claim to have found a narrow $\Delta^0$ bound nucleus and a
theoretical calculation by Walcher \cite{walcher} tried even to explain its
existence, these works were criticized in \cite{ramos} due to the importance
of the non-mesonic $\Delta$ decay, namely, $\Delta N \to N N$
(they found the width to be around 100 MeV) and the idea in
general remained mostly ignored.

Coming back to the discussion of the N$^*$-nuclei, 
in the present work we shall investigate the possibility for the existence of
N*-$^3$He and N*-$^{24}$Mg unstable bound states within a one meson
exchange model for the elementary N N* interaction. Though not very obvious, 
these states could possibly be 
related to the formation of $\eta$-$^4$He and $\eta$-$^{25}$Mg quasibound nuclei.
In the next section we present the the N$^*$-nucleus 
potentials and the method to locate the possible bound states of this potential. 

\section{N$^*$-nucleus potential}
Since the N$^*$-N interaction is not well known and the existence of
such a baryon resonance-nuclear state is as such not really known, in the present work 
we will try to make a simple estimate to see if any further sophisticated
calculation is worth following. 
With this in mind, we shall use (a) a
one meson exchange N N$^* \to$ N N$^*$ interaction
which is scalar and does not involve the spin dependent parts and (b) the
N$^*$-nucleus potential which is obtained by folding the elementary N N$^*$ interaction 
with a nuclear density. Neglecting the spin dependent parts is not a drastic assumption as
we will see below. Since the N$^*$(1535) is a negative parity baryon, indeed in the
one - pion and -eta exchange diagrams, the spin dependent terms are suppressed as
compared to the leading scalar terms.

\subsection{Elementary N N$^*$ $\to$ N N$^*$ potential}
The diagrams which we shall consider are shown in Fig. 1. We consider an N$^*$
which is neutral. The calculation for a positively charged N$^*$ can be repeated in
a similar way.
\begin{figure}[h]
\begin{center}
\includegraphics[width=9cm,height=3cm]{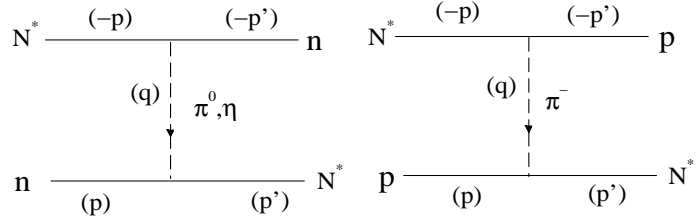}
\caption{\label{fig:eps1} Elementary N N$^*$ $\to$ N N$^*$ processes considered 
in this work }
\end{center}
\end{figure}
We shall not consider diagrams involving 
the N$^*$N$^*\,\pi$ or N$^*$N$^* \,\eta$ couplings which are hardly known. 
Apart from this fact, for such diagrams, the potential turns out to
be spin dependent (and so also suppressed as compared to the leading term in the 
potential of Fig. 1 ).

The $\pi$NN$^*$ and $\eta$NN$^*$ couplings (with N$^*$(1535,1/2$^-$)) are given by
the following interaction Hamiltonians \cite{osetetaNN}:
\begin{eqnarray}\label{hamil}
\delta H_{\pi N N^*} = g_{\pi N N^*} \bar{\Psi}_{N^*}  {\vec \tau} \Psi_N \cdot 
{\vec \Phi_{\pi}} + {\rm h.c.}\\ \nonumber
\delta H_{\eta N N^*} = g_{\eta N N^*} \bar{\Psi}_{N^*}  \Psi_N \cdot 
\Phi_{\eta} + {\rm h.c.}
\end{eqnarray}
Starting with say the diagram for the N$^*$ n $\to$ n N$^*$
process in Fig. 1 and using the standard Feynman diagram rules with the
non-relativistic approximation for the spinors
\begin{equation}\label{spinoreq3}
u_i =\sqrt{2m_i}\left(\begin{array}{c} w_i\\
{\vec{\sigma}_i \cdot \vec{p}_i \over 2m_ic}\, w_i 
\end{array}\right) \, ,
\end{equation}
we can write the amplitude as
\begin{equation}
{g_{xNN^*}^2 \bar{u}_{N^*}(\vec{p}^{\, \prime}) \, u_n(\vec{p}) \,
\bar{u}_n(-\vec{p}^{\, \prime})\, u_{N^*}(-\vec{p}) \over q^2 - m_x^2}\, ,
\end{equation}
where $x = \pi$ or $\eta$ and $q^2= \omega^2 - \vec{q}^2$ is the four momentum 
squared carried by the exchanged meson ($q = p^{\prime} - p$ as shown in the figure). 
Here for example,
\begin{equation} 
\bar{u}_n(-\vec{p}^{\, \prime})\, u_{N^*}(-\vec{p}) = N \, \biggl( 1 \, -\, 
{\vec{\sigma}_n \cdot \vec{p}^{\,\prime} \vec{\sigma}_{N^*} \cdot \vec{p} \over 
4 m_N m_N^* c^2} \biggr ) 
\end{equation}
and we drop the second term in the brackets which is spin dependent as well as
$1/c^2$ suppressed.
The potential in momentum space obtained from the above amplitude is given as:
\begin{equation}\label{pot1}
v_x(q) = {g^2_{xNN^*} \over q^2 - m_x^2} \, 
\biggl ({\Lambda^2_x - m_x^2 \over \Lambda_x^2 - q^2} \biggr )^2\, , 
\end{equation}
where the last term in brackets has been introduced to take into account the
off-shellness of the exchanged meson. The momentum transfer 
$q^2 = \omega^2 - \vec{q}^2$ in the present calculation is approximated simply as 
$q^2 \simeq - \vec{q}^2$. The neglect of the energy transfer in the 
elastic N N$^*\, \to$ N N$^*$ process is not necessarily justified but introducing 
a finite energy transfer gives rise to poles in (\ref{pot1}) thus making the 
calculation of the N$^*$ nucleus potential a formidable task. Hence, restricting 
ourselves to a calculation within this approximation, we Fourier transform the potential
in (\ref{pot1}) to obtain the potential in $r$-space. 
The Fourier transform of (\ref{pot1}) can be calculated analytically and we get,
\begin{equation}
v_x(r) = {g^2_{xNN^*} \over 4 \pi} \,\biggl [ {1\over r} \biggl 
( e^{-\Lambda_x r} - e^{-m_x r} \biggr ) + {\Lambda_x^2 - m_x^2 \over 2 \Lambda_x} \, 
e^{-\Lambda_x r} \biggr ]\, .
\end{equation}

The elementary potentials for two different parameter sets of the coupling 
constants for the $\pi N N^*$ and $\eta N N^*$ vertices are shown in Fig. 2.

\subsection{{\rm N}$^*$- $^3${\rm He} and {\rm N}$^*$ - $^{24}${\rm Mg} potentials}
Once the elementary potential has been defined, we use the folding model
\begin{equation}
V(R) = \int \, d^3r\, \rho(r) \, v(|\vec{r} - \vec{R}|) \, , 
\end{equation}
to construct the N$^*$ nucleus potential $V(R)$ and write,
\begin{eqnarray}
V(R) &=& V_p(R) + V_n(R) \nonumber \\ 
&=& Z \,\int \, d^3r\, \rho_p(r) \, v_p(|\vec{r} - \vec{R}|) \, +\, 
N \,\int \, d^3r\, \rho_n(r) \, v_n(|\vec{r} - \vec{R}|) \,, 
\end{eqnarray}
where, $Z$ and $N$ are the number of protons and neutrons,
$v_n(r) = v_{\pi^0}(r) + v_{\eta}(r)$ and
due to the isospin factor appearing in the $\pi^-$ exchange diagram
(see Fig. 1 and Eq.(\ref{hamil})),
$v_p(r) =  v_{\pi^-}(r) \vec{\tau}_1 \cdot \vec{\tau}_2$.
We also assume $\rho(r) = \rho_n(r) = \rho_p(r)$ with $\rho(r)$ normalized
to 1.
After performing the angle integration, the above integral reduces
for example to
\begin{eqnarray}
V_n(R) = {-2 \pi A \over R} \, \int \, \biggl \{ 
{e^{-m_x (|r - R|)} - e^{-m_x (r + R)} \over m_x} \, - \, 
{e^{-\Lambda_x (|r - R|)} - e^{-\Lambda_x (r + R)} \over \Lambda_x} 
\nonumber \\
+ B \biggl [ \, \biggl({r+R \over \Lambda_x} +{1 \over \Lambda_x^2} \biggr ) 
\,e^{-\Lambda_x(r+R)}\, -\, \biggl ( {|r-R| \over \Lambda_x} + {1 \over 
\Lambda_x^2} \biggr ) \,e^ {-\Lambda_x |r - R|}\, \biggr]  
\,\biggr \}\,
r \, dr\, \rho(r), \nonumber 
\end{eqnarray}
where $A = g^2_{xNN^*}/4\pi$ and $B= (\Lambda_x^2 - m_x^2)/2\Lambda_x$.

In case of the $^3$He nucleus, the nuclear density $\rho(r)$ is a sum of
Gaussians \cite{mccarthy} and the above integral can in principle be
done analytically. 
However, such an attempt leads to lengthy expressions with error functions
and exponentials which are not particularly
enlightening and hence we rather perform the integral numerically.
The density for $^3$He is taken from \cite{mccarthy} and that for $^{24}$Mg
is assumed to have a standard Woods-Saxon form.
The N$^*$ nuclear potentials (in Fig. 2) can be fitted reasonably well to Woods Saxon
forms of potentials. This facilitates the search for bound states of this
potential.
\begin{figure}[h]
\includegraphics[width=5cm,height=8cm]{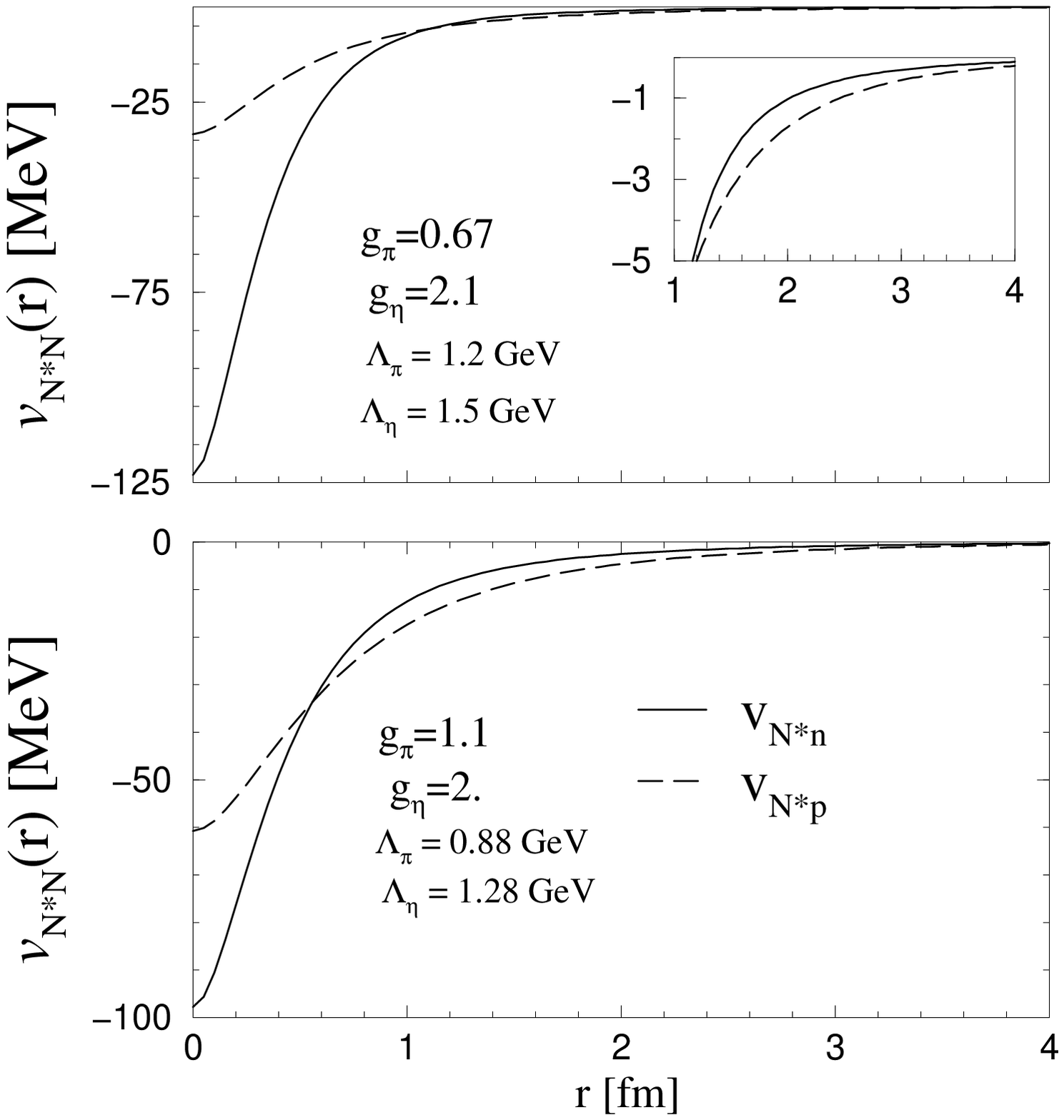}
\hskip 5mm
\includegraphics[width=5cm,height=8cm]{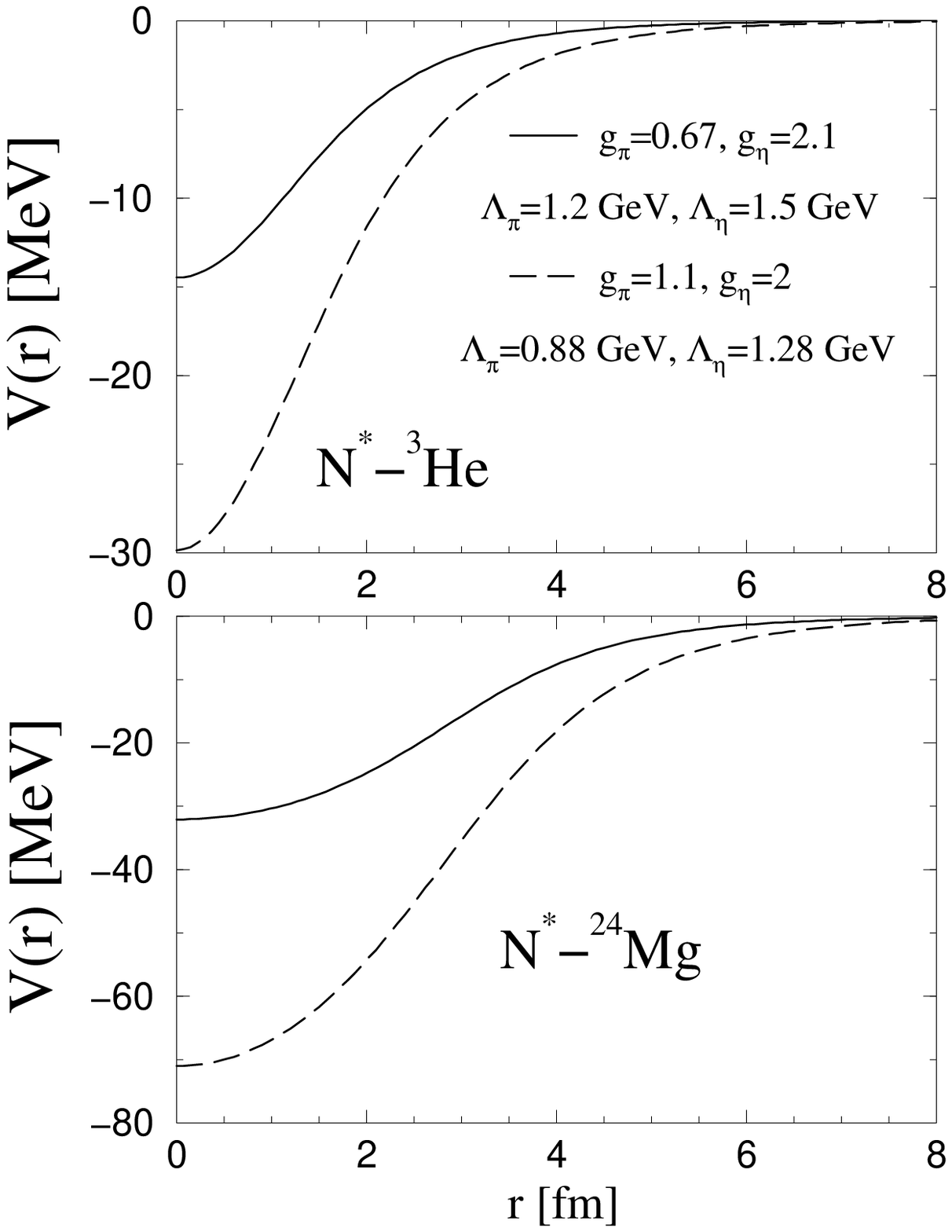}
\caption{\label{fig:fig1} Elementary N N$^*$ $\to$ N N$^*$ potentials with $\pi$ 
and $\eta$ exchange (left) and N$^*$-nuclear potentials (right). }
\end{figure}
\section{Bound states of the N$^*$-nucleus potential}
The Schr\"odinger equation for the Woods Saxon potential can be reduced to
one for the hypergeometric functions \cite{WShypergm} and a condition for
the existence of bound states can be found.
For a Woods Saxon potential of the type
\begin{equation}
V(r) = - {V_0 \over 1 + e^{r-R\over a}}
\end{equation}
the Schr\"odinger equation
\begin{equation}
{d^2u \over dr^2} + {2 \over r} {du\over dr} + {2m\over \hbar^2} (E - V) u =0  
\end{equation}
may be transformed to the independent variable
$y = 1 / [1 + e^{r - R/ a}]$ to obtain a hypergeometric differential equation. 
After some lengthy algebra \cite{WShypergm} one obtains the following
condition for bound states:
\begin{equation}\label{boundcond}
{\lambda R \over a} \, +\, \Psi \,-\, 2 \phi \, - \arctan{\lambda \over \beta} 
\, =\, (2n - 1) {\pi \over 2}\, \, \, \, \, n = 0, \pm 1, \pm 2, ...
\end{equation}
where,
$${2 m E \over \hbar^2} \, a^2 = - \beta^2 ; \,\,\, {2 m V_0 \over \hbar^2} \, 
a^2 =  \gamma^2 ; \, \, \, \lambda = \sqrt{\gamma^2 - \beta^2}$$
and $\phi = arg \Gamma (\beta + i \lambda)$; $\Psi = arg \Gamma (2 i \lambda)$.
\subsection{Binding energies of the {\rm N}$^*$-nuclei}
The N$^*$-nucleus potentials shown in Fig.2 can be very well fitted using a Woods 
Saxon form and the condition in (\ref{boundcond}) used to determine if a bound 
state of the N$^*$-nucleus potential exists and at what energy.
The table below gives the parameters of the Woods Saxon fits to 
the N$^*$-nucleus potentials and the energies of the bound states obtained 
using (\ref{boundcond}). The results 
are tabulated for two parameter sets of the coupling constants for the 
$\pi N N^*$ and $\eta N N^*$ vertices.  
\\
\\
\begin{tabular}{|l|l|l|}
  \hline
   & g$_{\pi N N^*}$ = 0.67, g$_{\eta N N^*}$ = 2.1  & 
g$_{\pi N N^*}$ = 1.1, g$_{\eta N N^*}$ = 2 \\
 & $\Lambda_{\pi}$ = 1.2, $\Lambda_{\eta}$ = 1.5 GeV \cite{BKSantra} &
$\Lambda_{\pi}$ = 0.88, $\Lambda_{\eta}$ = 1.28 GeV\\

   & $E$, $V_0$ [MeV], a, R [fm]  &  \cite{FixA}\\
  \hline
  N$^*$-$^3$He & $E$ = -0.03  & $E$ = -3.9  \\
      &  $V_0$=18, a=0.8, R=1.3  & $V_0$ = 37, a = 0.84, R = 1.4    \\
  \hline
  N$^*$-$^{24}$Mg   & $E$ = -17.1, -1.8 & $E$ = -47.6, -20.8, -2.6 \\
  & $V_0$ =34, a = 0.9, R = 2.9 & $V_0$ = 76, a = 0.98, R = 2.9\\
\hline
\end{tabular}

\subsection{Estimate of the widths} 
Given that the N$^*$-nucleus is not expected to be a ``bound" state (with 
infinite lifetime) but rather an unstable- or quasi-bound state, we also give a 
rough estimate of its width using a procedure similar to that of Ref.\cite{walcher}.
Assuming an average mean free path of the $\pi$ (or $\eta$) to be given by
$\langle l(\omega) \rangle = (\rho \, \sigma (\omega))^{-1}$ and also assuming
that the N$^*$ was produced say at the centre of the nucleus, the number of times
that the meson rescatters is given by
\begin{equation}
N(\omega) = g_{corr} \biggl ( {R \over \langle l(\omega) \rangle} \biggr )^2 
= g_{corr} [R \rho \sigma (\omega)]^2\, , 
\end{equation}
where we assume as in \cite{walcher}, 
that the geometric factor $g_{corr}$ is to be multiplied if
it is assumed that the N$^*$ is homogeneously produced over the nucleus.
Starting with the amplitude as a function of the energy $\omega$ as
\begin{equation}
G(\omega) = G_0 {\hbar \over \sqrt{2 \pi}} {- i \over (\omega - \omega_0 - \epsilon) + 
i (\Gamma/2)} 
\end{equation}
and taking into account that the meson does not propagate as a plane wave between
rescatters in the nucleus (after being produced and absorbed due to the N$^*$ decay),
$|G(\omega)|^2$ is found to be
\begin{equation}
|G(\omega)|^2 = G_0^2 {\hbar^2 \over 2 \pi} {1 \over (\omega - \omega_0 - \epsilon)^2 
+ (\Gamma /2)^2 } {\sin^2{((N(\omega) + 1) \phi(\omega)/2)} \over 
\sin^2{(\phi(\omega)/2)} } 
\end{equation}
where $\phi(\omega)$ is the phase advance experienced by the propagating meson and
is given by
\begin{equation}
\phi(\omega) = \arctan{\biggl ( {\omega_0 + \epsilon - \omega \over \Gamma/2 } \biggr )} 
\end{equation}
and
\begin{equation}
\sigma(\omega) = \sigma_0\, {(\Gamma/2)^2 \over (\omega - \omega_0 -\epsilon)^2 
+ (\Gamma/2)^2 } \, .
\end{equation}
Here $\omega_0$ is the difference of the N$^*$ and N masses ($\sim$ 597 MeV).
In Fig. 3 we see a plot of the function $|G(\omega)|^2$ (normalized to its peak value)
as a function of $\omega$ for different values of the cross section parameter $\sigma_0$.
The peak position is shifted from 597 MeV due to the meson phase factor as well as
the binding energy of the N$^*$ in the nucleus.
As we can see, the distributions become narrow for increasing values of the
cross sections. In the same figure, to the right, 
we see the full width at half maximum as a function
of $\sigma_0$ for the N$^*$-$^3$He and N$^*$-$^{24}$Mg nuclei. The absorption 
cross section parameter, 
$\sigma_0$ depends on the magnitude of the cross sections in $\pi N \to \pi N$
and $\eta N \to \eta N$ scattering in the N$^*$ resonance region. These cross sections
are of the order of 3 fm$^2$ for example for $\pi^- p \to \pi^- p + \pi^0 n$ in the
N$^*$ resonance region. In Fig. 3 we also see the maximum number of rescatters
that the meson would perform before leaving the nucleus at each 
value of $\sigma_0$. It appears from the figure that
for the range of relevant cross sections, the meson will not even rescatter once and
in this case the state would be broad (for example at $\sigma_0$ = 3 fm$^2$, 
$\Gamma \sim$ 80 MeV for N$^*$-$^3$He and 
about 110 MeV for N$^*$-$^{24}$Mg). 
It seems only consistent
that if the cross sections are bigger then there are more rescatters and the state is
longer lived (small $\Gamma$) as seen in the figure.
The curves in Fig. 3 are not very sensitive to the binding energy of the N$^*$-nucleus.
\begin{figure}[ht]
\includegraphics[width=6cm,height=6cm]{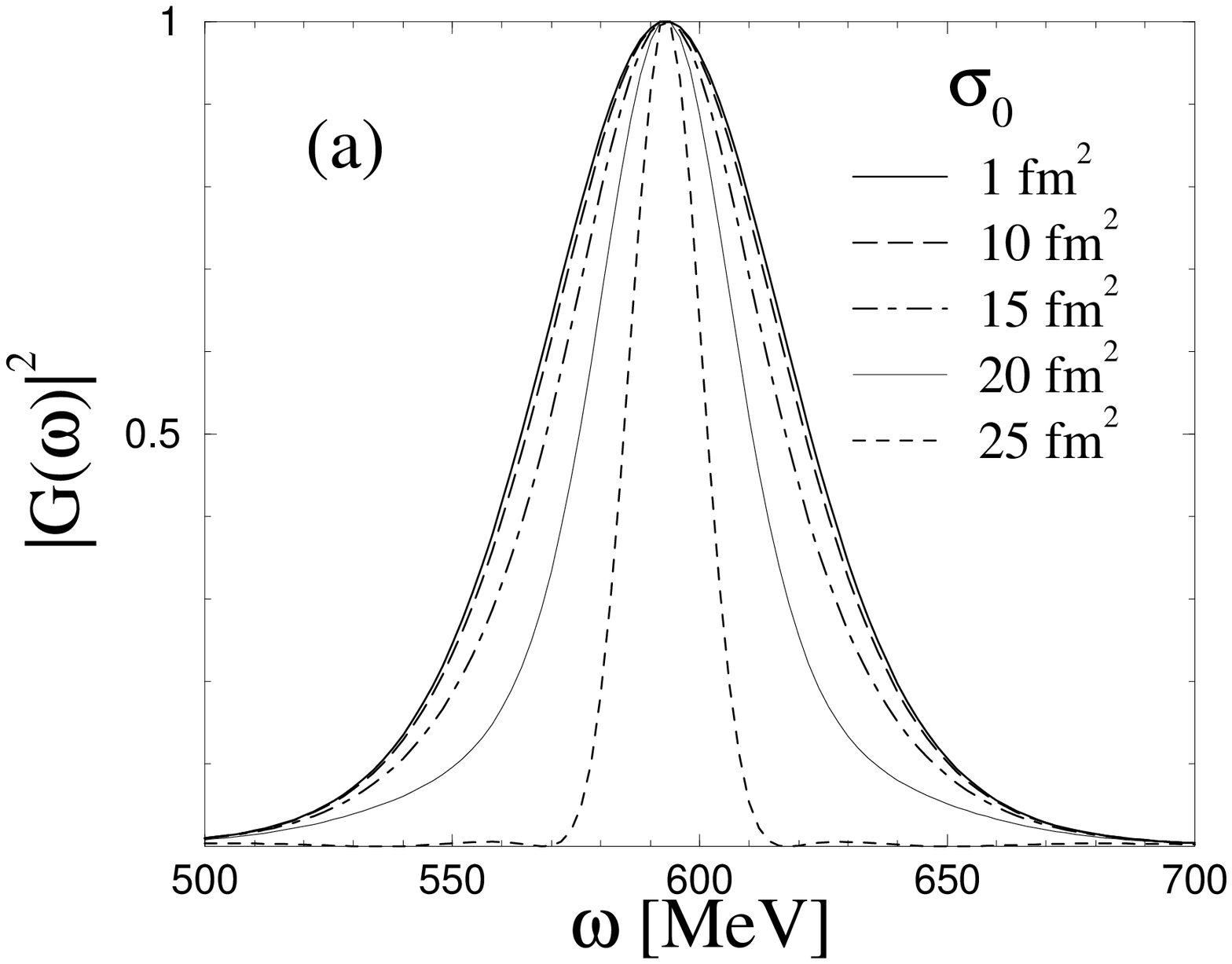}
\hskip 5mm
\includegraphics[width=6cm,height=6cm]{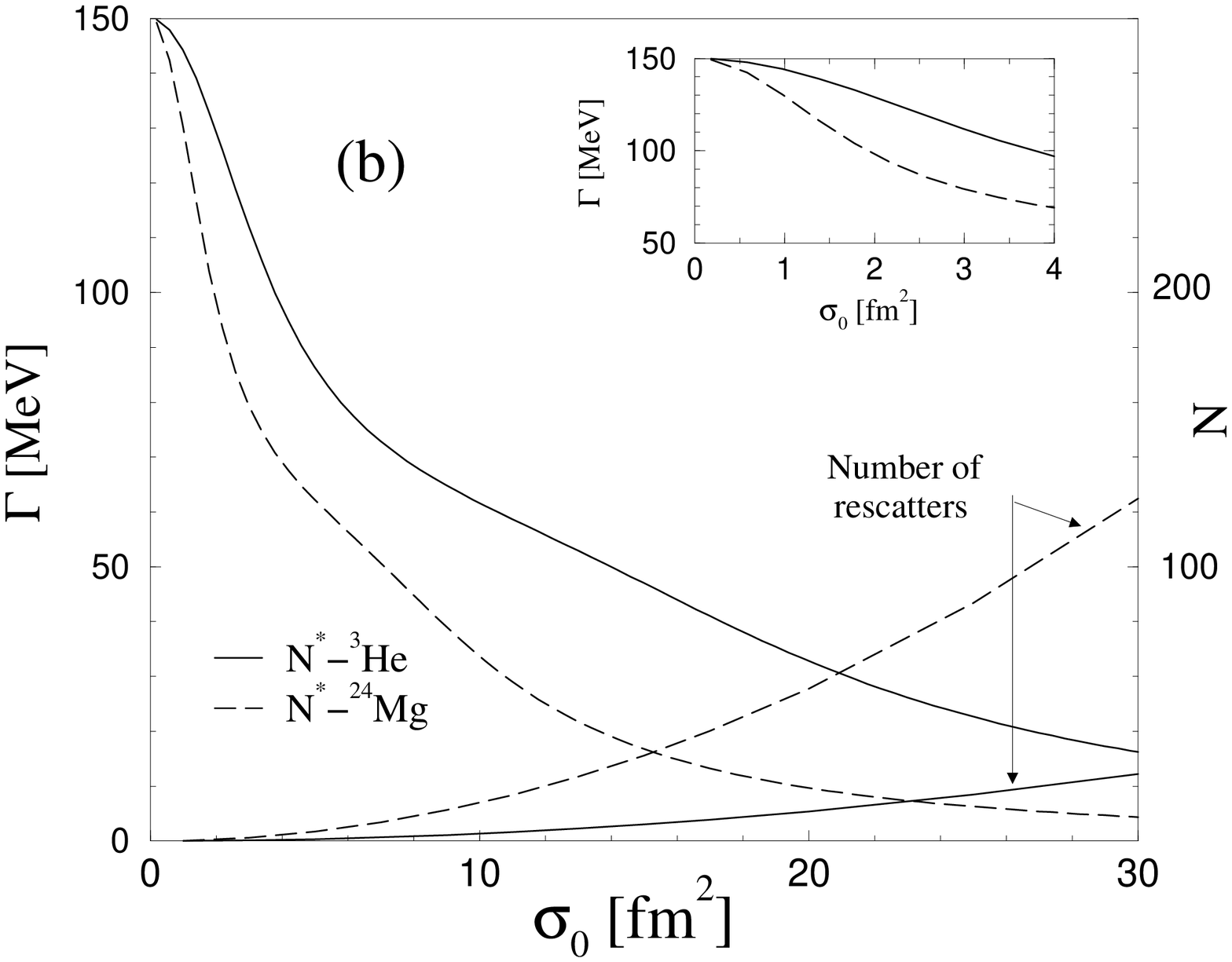}
\caption{\label{fig:epsfwhm} (a) Energy dependence of the normalized resonance curve 
$|G(\omega)|^2$ for different values of $\sigma_0$ and (b) the full width at 
half maximum ($\Gamma$) of the resonance curves as a function of $\sigma_0$. The maximum 
number of rescatterings (N) of the exchanged meson are also shown as a function of 
$\sigma_0$ (with the scale on the right side). }
\end{figure}

To summarize, we can say that within the simple model 
calculation done here, very broad states of N$^*$-$^3$He and N$^*$-$^{24}$Mg may 
exist. If an eta-mesic nucleus is visualized in the form of an eta meson propagating 
inside the nucleus via the formation, decay and regeneration of the N$^*$ resonance, 
the above could imply the existence of broad eta-mesic nuclear states \cite{broad}.


\begin{thebibliography}{99}
\bibitem{etamesic1}
W. Krzemien, P. Moskal and M. Skurzok, {\it Acta Phys. Polon. B} {\bf 46} 757 (2015); 
P. Adlarson {\it et al.}, Phys. Rev. C {\bf 87}, 035204 (2013); A. Budzanowski  
{\it et al}., Phys. Rev. C {\bf 79} 012201 (2009); 
M. Skurzok, Ph.D. thesis, Jagiellonian University (2015), arXiv:1509.01385.
\bibitem{etamesic2}
Q. Haider and L. C. Liu, ``Eta-mesic nuclei: past, present, future", arXiv: 1509.05487; 
H. Machner, {\it J.Phys. G} {\bf 42} 043001 (2015); 
B. Krusche, C. Wilkin, {\it Prog. Part. Nucl. Phys.} {\bf 80}, 43 (2014); 
N. G. Kelkar, K. P. Khemchandani, N. J. Upadhyay and B. K. Jain, 
{\it Rep. Prog. Phys.} {\bf 76}, 066301 (2013).
\bibitem{bartsch}
P. Bartsch {\it et al}., Eur. Phys. J. A {\bf 4}, 209 (1999).
\bibitem{walcher}
T. Walcher, Phys. Rev. C {\bf 63}, 064605 (2001).
\bibitem{ramos}
C. Chumillas, A. Parre\~no and A. Ramos, Nucl. Phys. A {\bf 791}, 329 (2007).
\bibitem{hirata}
M. Hirata, F. Lenz and K. Yazaki, Annals of Physics {\bf 108}, 116 (1977).
\bibitem{osetetaNN}
B. Lopez Alvaredo and E. Oset, Phys. Lett. B {\bf 324}, 125 (1994).
\bibitem{mccarthy}
J. S. McCarthy, I. Sick and R. R. Whitney, Phys. Rev. C {\bf 15}, 1396 (1977).
\bibitem{WShypergm}
S. Fl\"ugge, {\it Practical Quantum Mechanics}, Springer (1998);
M. Ghominejad, Eur. Phys. J. Plus {\bf 128}, 59 (2013).
\bibitem{BKSantra}
A. B. Santra and B. K. Jain, Nucl. Phys. A {\bf 634}, 309 (1998).
\bibitem{FixA}
A. Fix and H. Arenh\"ovel, Nucl. Phys. A {\bf 697}, 277 (2002).
\bibitem{broad}
K. Tsushima {\it et al}., Phys. Lett. B {\bf 443}, 26 (1988); 
C. Garcia-Recio, T. Inoue, J. Nieves and E. Oset, Phys. Lett. B {\bf 550}, 47 (2002).
\end{thebibliography}
\end{document}